\documentclass[twocolumn]{aastex631}
\let\splitbox\relax

\usepackage{tabularx}
\usepackage{colortbl}
\usepackage{adjustbox}

\emergencystretch=\maxdimen
\shorttitle{O3 Lite IMBHs}
\shortauthors{Ruiz-Rocha, Yelikar et al.}

\graphicspath{{./}{figures/}}

\usepackage{amsmath}
\usepackage{bm}

\begin{document}

\title{Properties of `Lite' Intermediate-Mass Black Hole Candidates 
in LIGO-Virgo's Third Observing Run}

\author{Krystal Ruiz-Rocha}
\affiliation{Department Physics and Astronomy, Vanderbilt University, 2301 Vanderbilt Place, Nashville, TN, 37235, USA}

\author{Anjali B. Yelikar}
\affiliation{Department Physics and Astronomy, Vanderbilt University, 2301 Vanderbilt Place, Nashville, TN, 37235, USA}

\author{Jacob Lange}
\affiliation{INFN Sezione di Torino, Via P. Giuria 1, 10125 Torino, Italy}

\author{William Gabella}
\affiliation{Department Physics and Astronomy, Vanderbilt University, 2301 Vanderbilt Place, Nashville, TN, 37235, USA}

\author{Robert A. Weller}
\affiliation{Department Physics and Astronomy, Vanderbilt University, 2301 Vanderbilt Place, Nashville, TN, 37235, USA}

\author{Richard O’Shaughnessy}
\affiliation{Center for Computational Relativity and Gravitation, Rochester Institute of Technology, Rochester, New York 14623, USA}

\author{Kelly Holley-Bockelmann}
\affiliation{Department Physics and Astronomy, Vanderbilt University, 2301 Vanderbilt Place, Nashville, TN, 37235, USA}

\author{Karan Jani}
\affiliation{Department Physics and Astronomy, Vanderbilt University, 2301 Vanderbilt Place, Nashville, TN, 37235, USA}

\begin{abstract}
Over a hundred gravitational-wave (GW) detections and candidates have been reported  from the first three observing runs of the Advanced LIGO-Virgo-KAGRA (LVK) detectors. Among these, the most intriguing events are binary black hole mergers that result in a `lite' intermediate-mass black hole (IMBH) of ${\sim}10^2~\mathrm{M}_\odot$, such as GW170502 and GW190521.
In this study, we investigate 11 GW candidates from LVK’s Third Observing Run (April 2019–March 2020) that have a total detector-frame masses in the lite IMBH range. Using the Bayesian inference algorithm \texttt{RIFT}, we systematically analyze these candidates with three state-of-the-art waveform models that incorporate higher harmonics, which are crucial for resolving lite IMBHs in LVK data. For each candidate, we infer the pre-merger and post-merger black hole masses in the source frame, along with black hole spin projections across all three models.
Under the assumption that these are binary black hole mergers, our analysis finds that 5 of them have a post-merger lite IMBH with masses ranging from $110\sim 350~\mathrm{M}_\odot$ with over 90\% confidence interval. Additionally, we note that one of their pre-merger black holes is within the pair-instability supernova mass gap ($60-120~\mathrm{M}_\odot$) with more than 90\% confidence interval, and additional two pre-merger black holes above the mass-gap. Furthermore, we report discrepancies among the three waveform models in their mass and spin inferences of lite IMBHs, with at least three GW candidates showing deviations beyond accepted statistical limits.
While the astrophysical certainty of these candidates cannot be established, our study provides a foundation to probe the lite IMBH population that emerge within the low-frequency noise spectrum of LVK detectors.
\end{abstract}


\section{Introduction} \label{sec:intro}

While the origin stories of intermediate-mass black holes (IMBHs) in the universe remain unknown, it is broadly expected that astrophysical processes will produce this class of black holes in the mass range between $100~\mathrm{M}_\odot$ to $100,000~\mathrm{M}_\odot$~\citep{Greene_2020}. Finding IMBHs, therefore, provide a direct insight into the astrophysical processes that are hidden at the stellar and super-massive scales of black holes. Gravitational-wave (GW)  detectors on earth and in space could lead to unambiguous measurement of IMBHs across the entire mass range~\citep{2020NatAs...4..260J}, which has so far proved to be extremely challenging with electromagnetic (EM) surveys.

\begin{table*}[t!]
\centering
\begin{tabular}{|c|c|c|c|c|c|c|c|c|}
\hline
\multicolumn{1}{|c|}{{Event}} & \multicolumn{5}{|c|}{{Offline LVK Search Results}} & \multicolumn{3}{|c|}{{RIFT Parameter Estimation (This work)}}\\
\hline 
&FAR(yr$^{-1}$)&Pipeline&Detectors&SNR&Reported in& SEOBNR PHM & Phenom PHM & NRSur PHM\\
\hline
\textbf{GW190403\_051519}&7.7 & PyCBC& HLV & 8& GWTC-2.1& \checkmark& \checkmark& \checkmark\\ 
\textbf{GW190426\_190642}&4.1 & PyCBC & HLV & 9.6& GWTC-2.1& \checkmark& \checkmark& \checkmark\\ 
GW190924\_232654 & 0.4 & PyCBC & LV & 13.3 & O3 IMBH&\checkmark & \checkmark & \checkmark \\ 
GW191109\_010717& $1.8\times 10^{-4}$ & MBTA &HL  & 15.2& GWTC-3& \checkmark& \checkmark& \checkmark\\ 
GW191223\_014159 & 0.46 & GstLAL & HLV& 14.2& O3 IMBH&\checkmark & \checkmark & \checkmark \\
GW191225\_215715& $4.7\times 10^{-1}$ & PyCBC & LV & 19.7& O3 IMBH& \checkmark& \checkmark& \checkmark\\
GW191230$\_$180458 & $5\times10^{-2}$ & cWB & HLV & 10.3& GWTC-3& \checkmark& \checkmark& \checkmark\\
GW200114\_020818& $5.8\times10^{-2}$ & cWB  & HLV & 14.5& O3 IMBH& \checkmark& \checkmark& \checkmark\\ 
GW200214\_224526& $1.3\times10^{-1}$ & cWB & HL& 13.1& O3 IMBH& \checkmark& \checkmark& \checkmark\\
\textbf{GW200220\_061928} & 6.8 & PyCBC& HLV& 7.5 & GWTC-3& \checkmark & \checkmark & \checkmark\\
\textbf{GW200308\_173609}& 2.4 & PyCBC& HLV & 8 & GWTC-3& \checkmark & \checkmark & \checkmark \\
\hline
\end{tabular}
\caption{The events considered in the paper, offline Search results from CBC detection pipelines (PyCBC~\citep{DalCanton:2020vpm}, MBTA~\citep{Aubin:2020goo}, GstLAL~\citep{Tsukada:2023edh,Ewing:2023qqe} and cWB~\citep{Klimenko:2015ypf,Mishra:2022ott}), and the waveform analyses used for parameter estimation- SEOBNRv4PHM (SEOBNR PHM), IMRPhenomXPHM (Phenom PHM) and NRSur7dq4 (NRSur PHM). Events in bold fall under sub-category ii) mentioned in Section ~\ref{sec:methods}.}

\label{Events_table}
\end{table*}

All currently operating GW detectors on the Earth - Advanced LIGO~\citep{LIGOScientific:2014pky}, Advanced Virgo~\citep{VIRGO:2014yos} and KAGRA~\citep{KAGRA:2020tym}, have their maximum reach for merging binary black holes that produce a \textbf{`lite' IMBH} of a few $\sim10^2~\mathrm{M}_\odot$.  In the last few years, such lite IMBHs have been reported by the LIGO-Virgo-KAGRA (LVK) Collaboration, as well as by independent groups analyzing the public GW data~\citep{Nitz:2021zwj,Mishra:2024zzs}. Prominent examples have been GW190521~\citep{LIGOScientific:2020iuh,LIGOScientific:2020ufj}, GW190426\_190642~\citep{LIGOScientific:2021usb} and GW170502~\citep{Udall:2019wtd}, each of which have shown evidence of a ${\sim} 150~\mathrm{M}_\odot$ remnant black hole.

Lite IMBHs are arguably the most profound discoveries from GW astronomy thus far for two primary reasons. One, the component mass of their pre-merger black holes are within the (pulsational) pair-instability supernovae (PISN) mass-gap of $\sim50-130~\mathrm{M}_\odot$, a region where we do not expect any compact object from stellar evolution~\citep{Woosley:2021xba,Farag:2022jcc}, although the bounds of this gap are not well-defined~\citep{Farmer:2019jed,Marchant:2020haw,Fishbach:2020qag}. Secondly, the remnant black hole is a few orders of magnitude lighter than any IMBH from EM observations. This work consistently uses the prefix \textit{lite} to highlight that the IMBHs discussed here are a different sub-class than the \textit{typical} IMBHs defined by the broader astronomy and EM community.

In this study, we conduct a comprehensive parameter estimation of the eleven lite IMBH candidates reported in the Third Observing Run of LIGO-Virgo (April 2019 - March 2020). The term \textit{candidates} is an ad-hoc distinction, since some of these events are labeled with `GW' tags, while others are classified as marginal events with lower detection statistics. Nevertheless, the defining criteria is that each of the candidate has been found by a LVK Collaboration search algorithm in at least two GW detectors, and the binary has a total mass of in the detector frame of $\gtrsim 100 \ \mathrm{M}_{\odot}$. In this category of $\mathrm{M}_\mathrm{tot}^\mathrm{det}$ $\gtrsim 100\ \mathrm{M}_{\odot}$, we further select events based on whether it falls under either of the two categories: i) it was found by at least one of the IMBH-dedicated search campaigns in ~\citep{LIGOScientific:2021tfm} ii) it has a low signal-to-noise ratio (SNR $\leq 10$) and a high false-alarm-rate (FAR $\geq$ 1 yr$^{-1}$). 

Section \ref{sec:methods} lists the IMBH candidates analyzed in this work and the GW Bayesian infrastructure utilized for their parameter estimation. In Section \ref{sec:results}, we report the pre-merger and post-merger properties for these candidates and compare them with previously reported lite IMBH events. We also discuss the prospects of these candidates being in the PISN mass-gap and the remnant in the IMBH region. In doing so, we conduct a systematic investigation of waveform models in the recovery of parameters for lite IMBHs. Section \ref{sec:discussion} describes the new insights we gain from these lite IMBH candidates on waveform systematics and low frequency detector noise.

\section{Methodology} \label{sec:methods}

\subsection{Lite IMBH Candidates}

Table ~\ref{Events_table} summarizes the event names considered for analysis in this paper, along with the offline CBC detection pipeline results (LVK Search results). GW190403\_051519 (referred to as GW190403) and GW190426\_190642 (referred to as GW190426), are significant candidates reported in GWTC-2.1~\citep{LIGOScientific:2021usb}. GW191109\_010717 (referred to as GW191109), GW191230\_180458 (referred to as GW191230), GW200220\_061928 (referred to as GW200220) and GW200308\_173609 (referred to as GW200308) are significant candidates reported in GWTC-3~\citep{KAGRA:2021vkt}. GW191109 is also a significant candidate found by the dedicated IMBH search. GW200114\_020818 (referred to as GW200114), GW200214\_224526 (referred to as GW200214), GW191225\_215715 (referred to as GW191225), GW190924\_232654 (referred to as GW190924) and GW191223\_014159 (referred to as GW191223) are marginally significant candidates reported in the O3 IMBH Search paper~\citep{LIGOScientific:2021tfm}. We use the same strain data and noise power spectral density(PSD) estimates as in the GWTC-2.1 and GWTC-3 catalog papers for common events, and use the deglitched-frames when available for an event. For signals that have not been previously analyzed, we estimate the detector strain
power spectrum (PSD) using  \texttt{BayesWave}~\citep{2015CQGra..32m5012C,PhysRevD.91.084034,Cornish:2020dwh}, consistent with the methods used to
generate the PSDs provided and used by the LVK to analyze the other events. 
We note that although all the candidates mentioned in this work are public, the parameter estimation results were not public for all of them. To the best of our knowledge, this work is the first to publish PE results for candidates from the O3 IMBH Search paper- GW190924, GW191223, GW191225, and GW200214. We perform a consistent study of analysis settings and samplers, with the only change being in waveform model assumption- to get clarity on any systematics.

\begin{table*}[ht!]
\centering
\tablewidth{0pt}
\renewcommand{\arraystretch}{1.5} 
\setlength{\tabcolsep}{2pt} 
\begin{tabular}{|l||c|c|c|c|c|c|c|c|c|c|}
\hline\hline
Event & M$_{\mathrm{tot}}^{\mathrm{det}}[\mathrm{M}_{\odot}]$ & M$_{\mathrm{tot}}^{\mathrm{src}}[\mathrm{M}_{\odot}]$ & $\mathrm{m}_1^{\mathrm{src}}[\mathrm{M}_{\odot}]$ & $\mathrm{m}_2^{\mathrm{src}}[\mathrm{M}_{\odot}]$ & $\chi_\mathrm{eff}$ & $\chi_\mathrm{p}$ & $\mathrm{M}_{\mathrm{f}}^{\mathrm{src}}[\mathrm{M}_{\odot}]$ & $\chi_{\mathrm{f}}$ & D$_\mathrm{L}$[Gpc] & z\\
\hline
GW190403 & $219^{+26}_{-27}$ & $87^{+29}_{-21}$ & $58^{+24}_{-19}$ & $28^{+17}_{-9}$ & $0.50^{+0.17}_{-0.23}$ & $0.32^{+0.26}_{-0.20}$ & $82^{+28}_{-20}$ & $0.84^{+0.04}_{-0.07}$ & $11.4^{+7.6}_{-5.5}$ & $1.52^{+0.78}_{-0.63}$\\
GW190426 & $299^{+43}_{-31}$ & $167^{+43}_{-34}$ & $96^{+35}_{-23}$ & $70^{+24}_{-25}$ & $0.07^{+0.39}_{-0.25}$ & $0.43^{+0.30}_{-0.30}$ & $159^{+41}_{-33}$ & $0.71^{+0.12}_{-0.11}$ & $5.1^{+3.9}_{-2.7}$ & $0.79^{+0.47}_{-0.36}$\\
GW190924 & $374^{+64}_{-72}$ & $201^{+56}_{-62}$ & $120^{+48}_{-39}$ & $79^{+35}_{-34}$ & $0.48^{+0.29}_{-0.46}$ & $0.45^{+0.34}_{-0.29}$ & $189^{+52}_{-58}$ & $0.85^{+0.08}_{-0.17}$ & $5.6^{+6.9}_{-3}$ & $0.86^{+0.79}_{-0.40}$\\
GW191109 & $136^{+13}_{-13}$ & $103^{+15}_{-13}$ & $62^{+11}_{-13}$ & $41^{+11}_{-10}$ & $-0.37^{+0.24}_{-0.21}$ & $0.61^{+0.28}_{-0.36}$ & $99^{+14}_{-13}$ & $0.56^{+0.12}_{-0.15}$ & $1.6^{+1.6}_{-0.8}$ & $0.30^{+0.24}_{-0.14}$\\
GW191223 & $468^{+73}_{-95}$ & $347^{+71}_{-86}$ & $262^{+55}_{-69}$ & $84^{+30}_{-21}$ & $-0.06^{+0.23}_{-0.38}$ & $0.27^{+0.55}_{-0.21}$ & $338^{+68}_{-83}$ & $0.54^{+0.19}_{-0.17}$ & $1.9^{+1.3}_{-0.8}$ & $0.34^{+0.20}_{-0.13}$\\
GW191225 & $338^{+37}_{-33}$ & $292^{+33}_{-31}$ & $229^{+26}_{-27}$ & $62^{+14}_{-7}$ & $-0.37^{+0.20}_{-0.16}$ & $0.49^{+0.19}_{-0.26}$ & $286^{+32}_{-31}$ & $0.40^{+0.14}_{-0.15}$ & $0.76^{+0.49}_{-0.31}$ & $0.16^{+0.09}_{-0.06}$\\
GW191230 & $145^{+18}_{-17}$ & $82^{+16}_{-11}$ & $47^{+12}_{-8}$ & $35.7^{+9.5}_{-9.3}$ & $-0.03^{+0.22}_{-0.25}$ & $0.38^{+0.33}_{-0.28}$ & $78^{+16}_{-10}$ & $0.68^{+0.08}_{-0.10}$ & $4.9^{+2.2}_{-2.2}$ & $0.77^{+0.27}_{-0.29}$\\
GW200114 & $163^{+25}_{-23}$ & $118^{+27}_{-20}$ & $75^{+30}_{-19}$ & $42^{+16}_{-13}$ & $-0.45^{+0.21}_{-0.18}$ & $0.37^{+0.24}_{-0.22}$ & $114^{+27}_{-19}$ & $0.49^{+0.11}_{-0.17}$ & $2^{+1.7}_{-1.1}$ & $0.37^{+0.24}_{-0.19}$\\
GW200214 & $272^{+51}_{-42}$ & $175^{+45}_{-32}$ & $116^{+32}_{-26}$ & $59^{+32}_{-24}$ & $0.07^{+0.35}_{-0.33}$ & $0.33^{+0.36}_{-0.25}$ & $168^{+42}_{-29}$ & $0.68^{+0.15}_{-0.17}$ & $3.4^{+1.4}_{-1.8}$ & $0.56^{+0.19}_{-0.28}$\\
GW200220 & $284^{+47}_{-41}$ & $129^{+43}_{-32}$ & $73^{+28}_{-19}$ & $56^{+22}_{-18}$ & $0.07^{+0.38}_{-0.35}$ & $0.46^{+0.39}_{-0.34}$ & $123^{+41}_{-30}$ & $0.72^{+0.13}_{-0.13}$ & $8.4^{+6.8}_{-4.3}$ & $1.19^{+0.74}_{-0.52}$\\
GW200308 & $90^{+230}_{-20}$ & $52^{+87}_{-14}$ & $37^{+58}_{-13}$ & $16^{+33}_{-6}$ & $0.50^{+0.27}_{-0.69}$ & $0.41^{+0.38}_{-0.27}$ & $49^{+85}_{-13}$ & $0.86^{+0.08}_{-0.29}$ & $6^{+12}_{-3}$ & $0.90^{+1.30}_{-0.40}$\\
\textbf{GW170502} & $364^{+70}_{-60}$ & $159^{+61}_{-44}$ & $96^{+45}_{-30}$ & $62^{+34}_{-24}$ & $0.28^{+0.35}_{-0.49}$ & $0.60^{+0.24}_{-0.32}$ & $151^{+58}_{-42}$ & $0.80^{+0.09}_{-0.17}$ & $9.3^{+8.2}_{-5.3}$ & $1.29^{+0.87}_{-0.64}$\\
\textbf{GW190521} & $273^{+26}_{-27}$ & $150^{+29}_{-17}$ & $85^{+21}_{-14}$ & $66^{+17}_{-18}$ & $0.08^{+0.27}_{-0.36}$ & $0.68^{+0.25}_{-0.37}$ & $142^{+28}_{-16}$ & $0.72^{+0.09}_{-0.12}$ & $5.3^{+2.4}_{-2.6}$ & $0.82^{+0.28}_{-0.34}$\\

\hline\hline
\hline\hline
\end{tabular}

\label{table:event summary}
\caption{Summary of parameters for the lite IMBHs analyzed in this study. Also listed are events GW170502~\citep{Udall:2019wtd} and GW190521~\citep{LIGOScientific:2020ufj} not analyzed in this work. We quote the median values of the parameters for the inference with the NRSur7dq4 model using \texttt{RIFT}, along with 90\% credible intervals. The columns show the event name, the detector-frame total mass M$^{\mathrm{det}}_{\mathrm{tot}}$,  the source-frame total mass M$^{\mathrm{src}}_{\mathrm{tot}}$, source-frame component masses m$^{\mathrm{src}}_{1}$ and m$^{\mathrm{src}}_{2}$, the effective aligned spin parameter $\chi_{\mathrm{eff}}$ and effective precession spin parameter $\chi_{\mathrm{p}}$, the remnant black hole properties: source-frame mass M$^{\mathrm{src}}_{\mathrm{f}}$ and spin magnitude $\chi_{\mathrm{f}}$, the luminosity distance D$_{\mathrm{L}}$ and corresponding redshift z.}

\end{table*}

\subsection{Bayesian inference}

A coalescing binary black hole in a quasi-circular orbit can be completely characterized by its intrinsic
and extrinsic parameters.  By intrinsic parameters, we refer to the binary's  (detector-frame) masses $m_i$ and spins $\chi_{i}$. The masses are expressed in terms of quantities chirp-mass ($\mathcal{M}={(m_{1}m_{2})^{3/5}}/{M^{1/5}}$) and mass ratio ($q={m_{2}}/{m_{1}}$) that better characterize the leading-order dependence of the gravitational wave phase than the individual masses.
By extrinsic parameters we refer to the seven numbers needed to characterize its spacetime location and orientation: luminosity distance ($\mathrm{D}_{\mathrm{L}}$), right ascension ($\alpha$), declination ($\delta$), inclination ($\iota$), polarization ($\psi$), coalescence phase ($\phi_{c}$), and time ($t_{c}$).
We will express masses in solar mass units and
 dimensionless spins in terms of Cartesian components $\chi_\mathrm{i,x},\chi_\mathrm{i,y}, \chi_\mathrm{i,z}$, expressed
relative to a frame with $\hat{\mathbf{z}}=\hat{\mathbf{L}}$ and (for simplicity) at the orbital frequency corresponding to the earliest
time of computational interest (e.g., an orbital frequency of $\simeq 10$~Hz). We will use \bm{$\lambda$} and \bm{$\theta$} to
refer to intrinsic and extrinsic parameters, respectively: 
$$ \bm{\lambda} : (\mathcal{M},q,\chi_\mathrm{1,x},\chi_\mathrm{1,y},\chi_\mathrm{1,z},\chi_\mathrm{2,x},\chi_\mathrm{2,y},\chi_\mathrm{2,z})$$ 
$$ \bm{\theta} : (\mathrm{D}_{\mathrm{L}},\alpha,\delta,\iota,\psi,\phi_{c},t_{c}) $$

In this work, we use the state-of-the-art Rapid Iterative FiTting (\texttt{RIFT}) algorithm \citep{gwastro-RIFT-Update} to determine the properties of lite IMBHs. This algorithm employs a Bayesian inference technique that operates through a two-stage iterative process to interpret GW observations \(d\) by comparing them to predicted GW signals \(h(\bm{\lambda}, \bm{\theta})\). At each iterative stage, for each \(\lambda_\beta\) within a proposed grid \(\beta=1,2,\ldots,N\) of candidate parameters (executed by multiple workers in parallel), \texttt{RIFT} computes a marginal likelihood: 

\begin{equation}    
{\cal L}{({\bm \lambda})} =\int  {\cal L}_{\rm full}(\bm{\lambda} ,\bm\theta )p(\bm\theta )d\bm\theta
\end{equation}
based on the full likelihood \({\cal L}_{\rm full}(\bm{\lambda} ,\theta )\) of the gravitational wave signal across a multi-detector network, accounting for detector responses. For further details, refer to \citep{gwastro-PE-AlternativeArchitectures, gwastro-PENR-RIFT, gwastro-RIFT-Update}.

In the second stage, \texttt{RIFT} performs two primary tasks. First, it generates an approximation of \({\cal L}(\lambda)\) based on its accumulated archive of marginal likelihood evaluations \((\lambda_\beta, {\cal L}_\beta)\). This approximation can be constructed using methods such as Gaussian processes, random forests, or other suitable techniques. Second, using this approximation, \texttt{RIFT} computes the (detector-frame) posterior distribution by performing a Monte Carlo integration:
\begin{equation}
  p_{\rm post}=\frac{{\cal L}(\bm{\lambda} )p(\bm{\lambda})}{\int d\bm{\lambda} {\cal L}(\bm{\lambda} ) p(\bm{\lambda} )}  
\end{equation}
where $p(\bm{\lambda})$ is the prior on intrinsic parameters like mass and spin.  
Thanks to optimizations tailored to evaluating the integrand and performing the marginal integral \({\cal L}_{\rm full}\) \citep{gwastro-PE-AlternativeArchitectures, gwastro-PE-AlternativeArchitecturesROM, gwastro-PENR-RIFT, gwastro-PENR-RIFT-GPU}, \texttt{RIFT}'s two-step architecture achieves a consistently low and duration-independent computational cost; see \citep{gwastro-RIFT-Update} for further details.

\begin{figure*}[hbtp]
    \includegraphics[width = 1\textwidth]{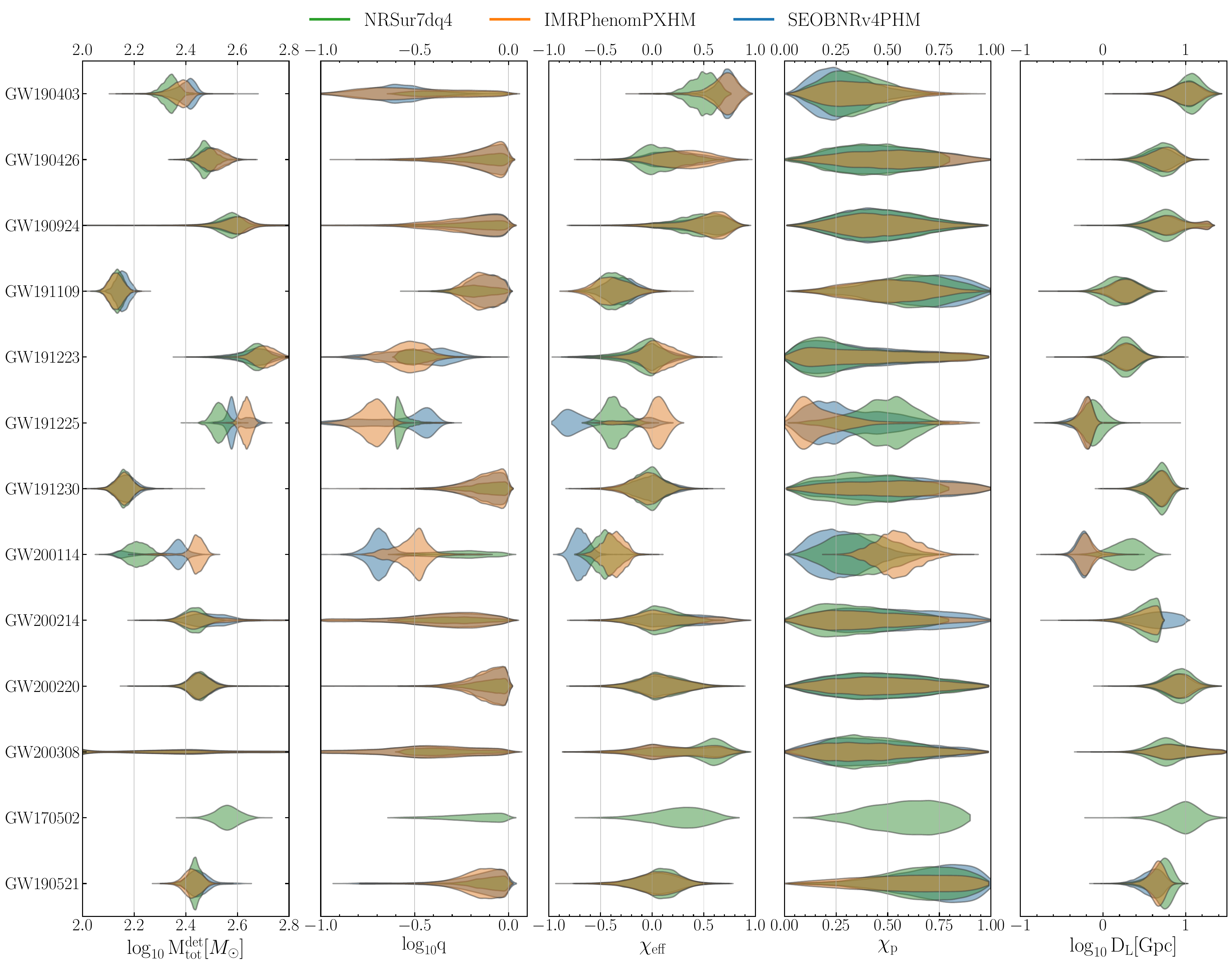}
    \caption{A comparison of analysis using different waveform models (NRSur7d4q, IMRPhenomXPHM, SEOBNRv4PHM) is displayed for each event. From left to right, we plot the detector-frame total mass ($\mathrm{M}_\mathrm{tot}^\mathrm{det}$), the mass ratio ($q$), the effective aligned spin parameter ($\chi_\mathrm{eff}$), the precession spin parameter ($\chi_\mathrm{p}$), and the luminosity distance ($\mathrm{D}_\mathrm{L}$). The $\mathrm{M}_\mathrm{tot}^\mathrm{det}$, $q$ and $\mathrm{D}_\mathrm{L}$ parameters are plotted as the natural logarithm of their respective values for visualization purposes. Note the GW190521 phenomenological model samples are from analysis using IMRPhenomPv3HM.}
    \label{figure:violin plots}
\end{figure*}

\begin{figure*}[t!]
    \centering
    \includegraphics[trim=0 5 0 0, clip, width = 0.9\textwidth]{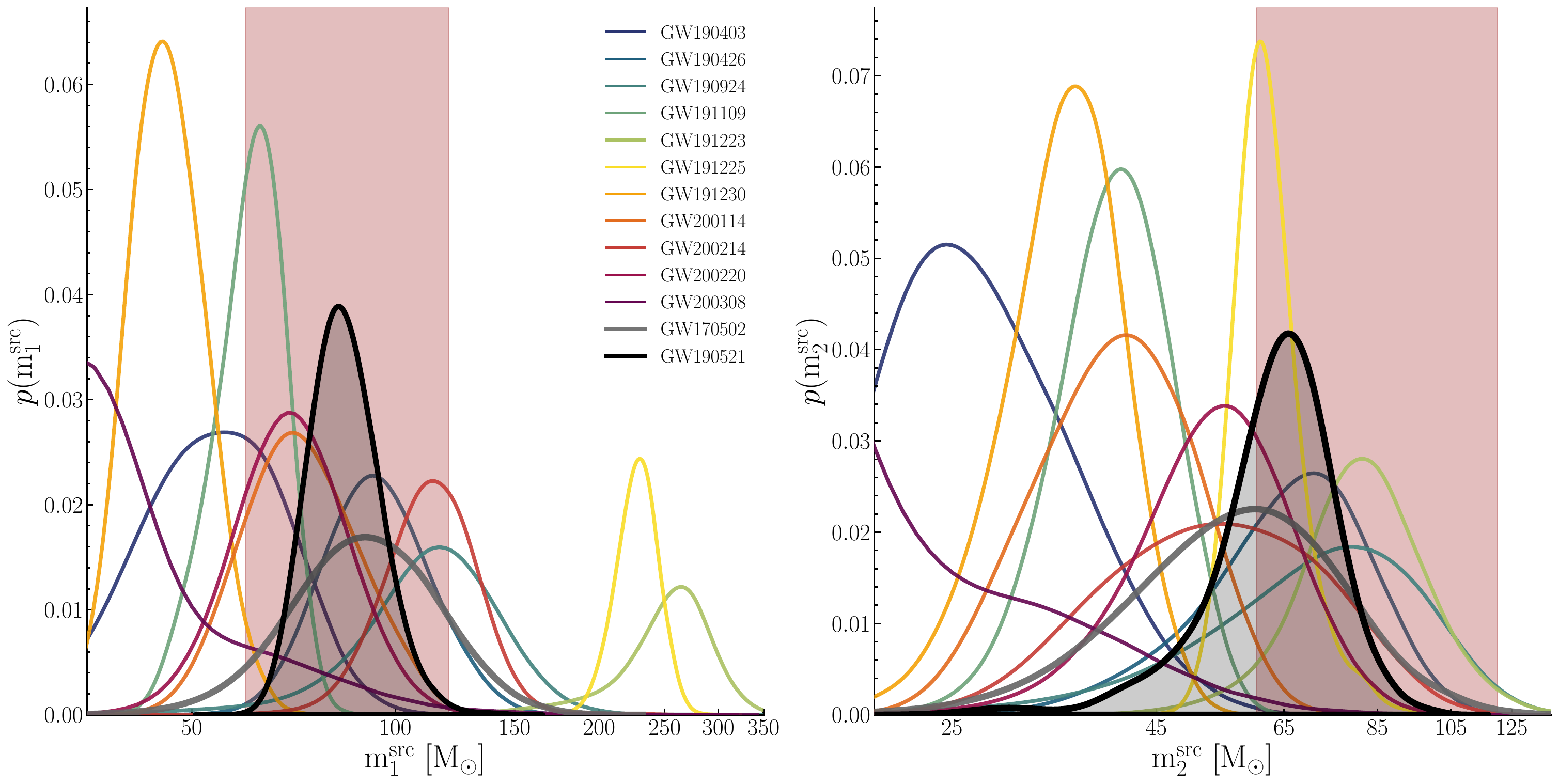}
    \caption{Marginal probability distribution of the primary \textit{(left)} and secondary \textit{(right)}  black holes in the source frame, as inferred through \texttt{RIFT} with NRSur7dq4 model. The grey and filled black posteriors are for GW170502 and GW190521, respectively, with NRSur7dq4 model. The red box highlights the expected PISN mass gap from [60-120]$M_\odot$.}
    
    \label{figure:mass_pdf}
\end{figure*}

\subsection{Waveform models and priors}

\texttt{RIFT} can perform inference using a variety of state-of-the-art GW approximation models for quasi-circular and eccentric binary black hole mergers. In this study, we use the most accurate available approximations that model precession across different waveform families as of the end of O3: the phenomenological model IMRPhenomXPHM \citep{Pratten:2020ceb}, the effective-one-body model SEOBNRv4PHM \citep{Ossokine:2020kjp}, and the numerical relativity surrogate model NRSur7dq4 \citep{Varma:2019csw}. For consistency, all three models are applied with a fixed setting, considering only modes up to \(\ell \leq 4\).

For the inference, we use a standard uniform prior over the detector-frame component masses and spin magnitudes, along with an isotropic prior for spin orientation, sky location, and binary orientation. The mass and mass-ratio limits are customized for each event to optimize computational time and align with the constraints of the waveform model in use (e.g., NRSur7dq4 is valid only for \(q \in [0.1667, 1]\) \citep{Varma:2019csw}).

The GW strain and noise power spectral density (PSD) for each event is sampled at a rate of 4kHz. Around the time of detection of every event, 4s of data are analyzed for GW190403, GW190426, GW190924, GW191109, GW191223, GW191225, GW191230 and GW200220, 8s of data is analyzed for GW200114 and GW200214, and 16s of data is analyzed for GW200308. 

We utilize the \texttt{PESummary} package \citep{Hoy:2020vys} to collate and post-process all the posterior samples in the cosmologically corrected source frame. To do so, we use the cosmological parameters from \texttt{Planck15}~\citep{Placnk15}. Our datasets are in the format consistent with those in the LVK gravitational wave catalog data releases \citep{LIGOScientific:2021usb}.

\section{Source Properties} \label{sec:results}

\begin{figure*}[ht!]
    \centering
      \includegraphics[trim= 10 0 150 0, clip, width= 0.42\textwidth]{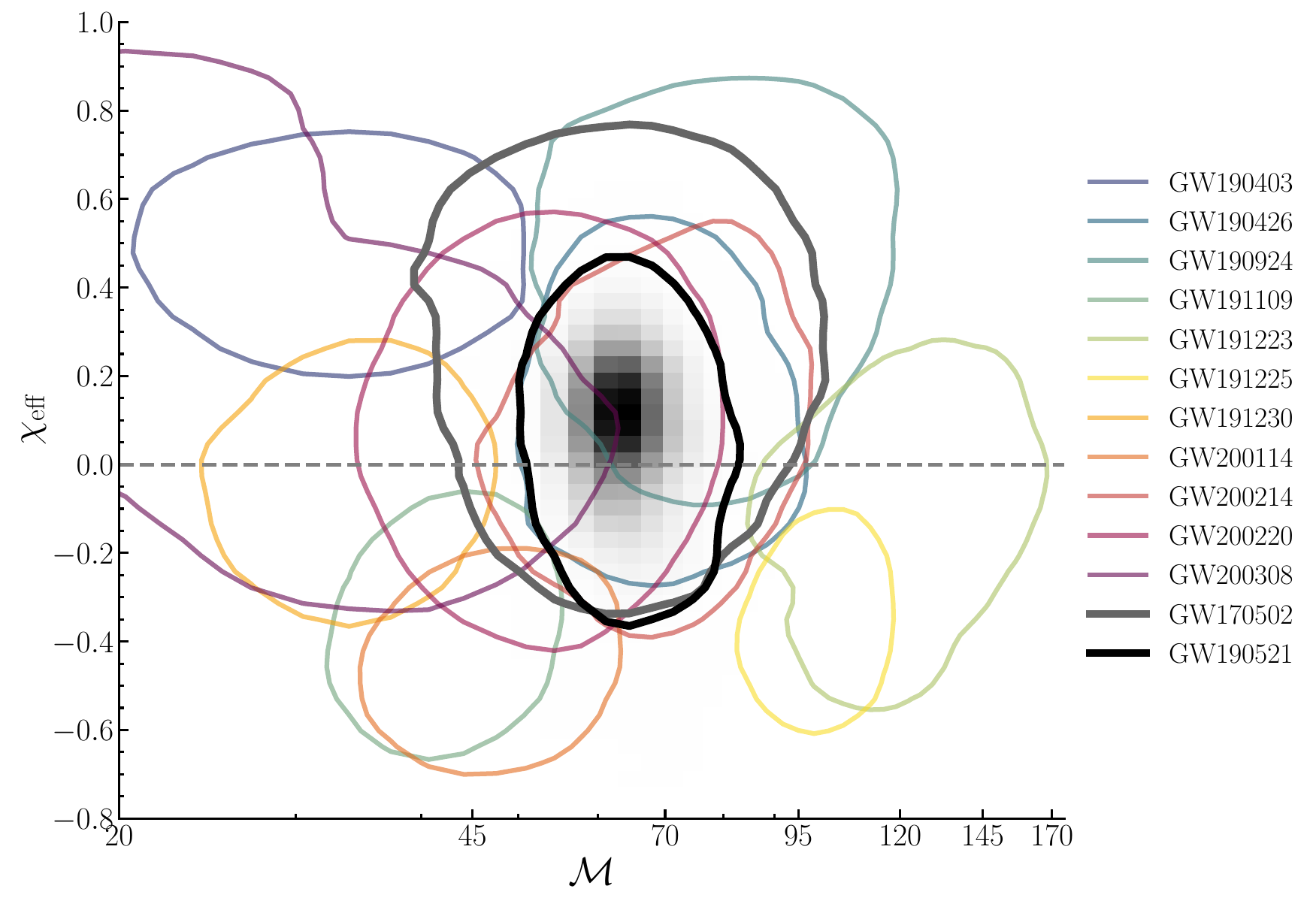}
    \includegraphics[width=0.5\textwidth]{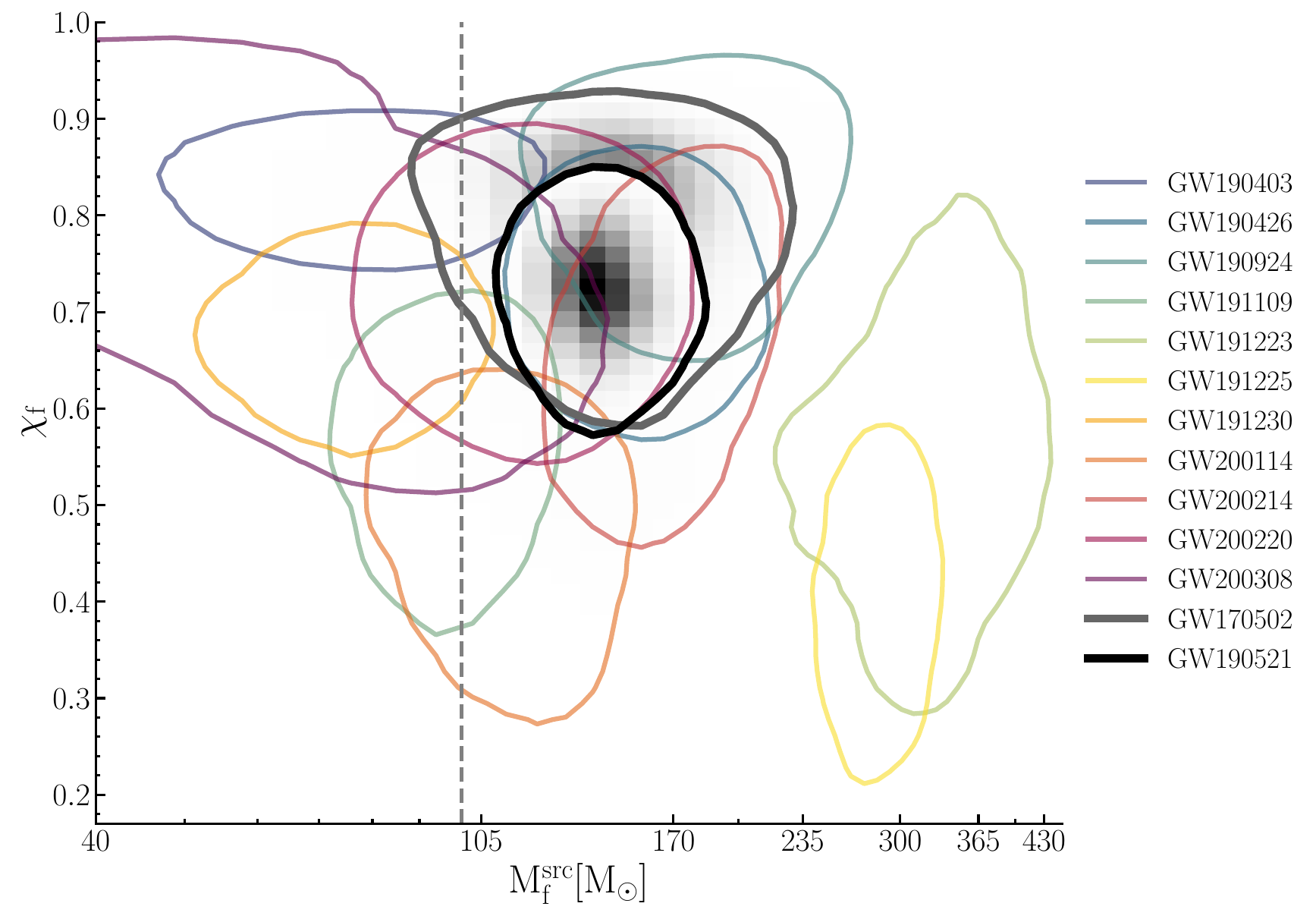}
    \caption{\textit{Left}: Posterior distribution on the chirp mass of the binary in source frame as a function of the inferred effective inspiral spin parameter. \textit{Right}: Posterior distributions of the mass and the dimensionless spin of the remnant black hole according to the RIFT inference using NRSur7dq4 model. The two-dimensional plot for both panels shows the 90\% credible regions of inference using \texttt{RIFT} with NRSur7dq4 model. The filled contour are the LVK posteriors of GW190521 with NRSur7dq4 model.}
         \label{figure:chirp mass chieff}
\end{figure*}%

In Table \ref{table:event summary}, we list the intrinsic and extrinsic parameters of the eleven lite IMBH candidates analyzed in this study. We quote the results from \texttt{RIFT} inference performed using NRSur7dq4 waveform model, following what was done for GW190521~\citep{LIGOScientific:2020iuh,LIGOScientific:2020ufj}. Unless stated otherwise, all the masses are quoted in the source frame after correcting for cosmological redshift corresponding to the candidate's inferred luminosity distance:
\begin{equation}
 m^\mathrm{src} = \frac{m^\mathrm{det} }{1+z(\mathrm{D}_\mathrm{L})}.   
\end{equation}

\subsection{Pre-merger black holes}

In Figure \ref{figure:violin plots}, we compare select intrinsic and extrinsic parameters of the events analysed in this work, GW170502 samples from~\citep{Udall:2019wtd} and GW190521 samples from~\citep{LIGOScientific:2020ufj} across the different waveform analysis. The first two columns show mass quantities: the total mass in the detector frame and the mass ratio. We ensure convergence of the respective analyses, and any differences in the inferred parameters from the waveform models can be owed to the differences in their modeling assumptions and limitations. The detector frame total mass ranges from 90.0$^{+230.0}_{-20.0}$ $\mathrm{M}_\odot$ to 468.0$^{+73.0}_{-95.0}$ $\mathrm{M}_\odot$. The inferred total mass distribution for GW200308 is bimodal and falls below our selection criterion of 100 $\mathrm{M}_\odot$, this could be due to the limitations of the NRSur7dq4 model that prefers a lower total mass. GW190426 with an $\mathrm{M}_\mathrm{tot}^\mathrm{src}$ of 167.0$^{+43.0}_{-34.0}$ $\mathrm{M}_\odot$ has close resemblance to the GW190521 and GW170502 masses. GW191225 stands out as the closest binary at 0.76$^{+0.49}_{-0.31}$ Gpc and relatively massive ($\mathrm{M}_\mathrm{tot}^\mathrm{src}$=292.0$^{+33.0}_{-31.0}$ $\mathrm{M}_\odot$). GW190403 with an inferred luminosity distance of 11.4$^{+7.6}_{-5.5}$ Gpc, is the most distant lite IMBH merger observed.

By decomposing the total source mass into primary and secondary mass, we plot the respective marginal probability density function for all events, including GW170502(gray) and GW190521(black) in Figure \ref{figure:mass_pdf}. The primary mass distributions exhibit peaks clustered between $50-130$ M$_\odot$, and the events notably fall within or near the predicted pair-instability supernova (PISN) mass gap. This is true for all events except for GW191223 and GW191225, which peak beyond 200 M$_\odot$. Events GW191230 and GW191109 show more tightly constrained primary mass distributions than the other events, surpassing the precision seen in GW190521. The secondary mass distributions(right panel) are less clustered than the primary mass. As depicted in the right panel, the mass distribution of secondary masses exhibits less clustering than the primary mass distribution. For these events, the peaks range from approximately 25 to 70 M$_\odot$, and it is noteworthy that about half of the events fall within the pair-instability supernova (PISN) mass gap between $[60-120]~\mathrm{M}_\odot$. Notably, the secondary mass distributions appear to be less constrained when compared to the primary mass distributions.

The third and fourth columns in Fig. \ref{figure:violin plots} compare the effective aligned spin ($\chi_\mathrm{eff}$) and precession spin ($\chi_\mathrm{p}$) parameters for all the events, including GW190521 using violin plots. The effective aligned spin ($\chi_\mathrm{eff}$) is a mass-weighted linear combinations of the individual spin components aligned with respect to the Newtonian orbital angular momentum ($\vec{L}$), and defined as:
\begin{equation}
\chi_\mathrm{eff} = \frac{m_{1}\chi_\mathrm{1,z} + m_{2}\chi_\mathrm{2,z}}{\mathrm{M}}    
\end{equation}

The effective precession spin parameter relates the spin components perpendicular to $\vec{L}$, and defined as: 
\begin{equation}
\chi_\mathrm{p} = \mathrm{max} \left( \chi_\mathrm{1,\perp},~q~\frac{4q+3}{4+3q}\chi_\mathrm{2,\perp} \right)    
\end{equation}

For the precession spin parameter ($\chi_\mathrm{p}$), we see the violin plots extending across the full range from 0 to 1.0, showing that the parameter is not highly constrained. We also examine 90\% credible region contours for effective aligned spin ($\chi_\mathrm{eff}$) and the chirp mass ($\mathcal{M}$) in left panel of Figure \ref{figure:chirp mass chieff}. For events similar to GW190521 and GW170502 in $\mathcal{M}$, they are clustered around an effective spin of 0.1 to 0.3.

\subsection{Post-merger black holes}

After merger, the final spins ($\chi_\mathrm{f}$) and final mass ($\mathrm{M}^\mathrm{src}_\mathrm{f}$) of the remnant are shown as 90\% contours in right panel of Figure \ref{figure:chirp mass chieff}. We obtain these remnant properties through post-processing of the pre-merger quantities using \texttt{PESummary}~\citep{Hoy:2020vys}, which averages over three numerical relativity (NR) fits ~\citep{Keitel:2016krm,Healy:2016lce,Hofmann:2016yih}. Similar to pre-merger properties, events GW191223 and GW191225 stand out from the rest. The other events have final spins between 0.4 to 0.8 and total masses between 50 and 330 $\mathrm{M}_\odot$. Of these events, GW200214, GW190426, GW191225, and GW200114 resulting in final masses in the IMBH mass range. Notably, GW191223 has the largest final mass, with GW191225 just below it, both being massive compared to GW190521.

\subsection{Astrophysical Implications}

\begin{figure}[b!]
    \centering
    \includegraphics[trim=45 30 0 0, clip,width = 1.07\linewidth]{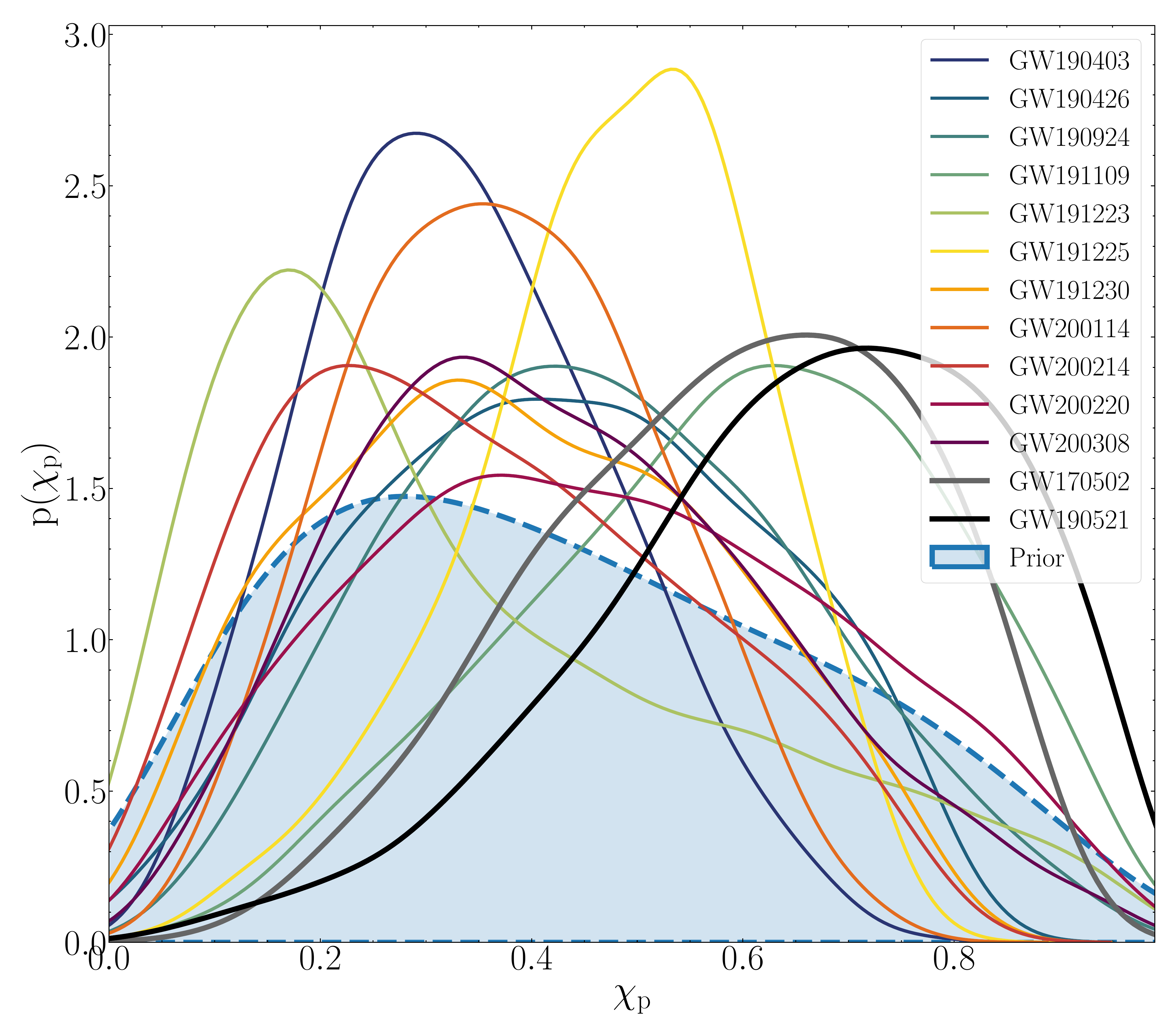}
    \caption{Marginal probability distribution of the effective precession spin parameter ($\chi_\mathrm{p}$) for the events as inferred through \texttt{RIFT} with NRSur7dq4 model, compared against the prior (dashed blue) on the parameter.}
    \label{fig:chi_p_pdf}
\end{figure}


The spin parameter of a system could help in understanding the formation channels of such binaries. We infer a negative $\chi_\mathrm{eff}$ for GW191109, GW191225 and GW200114, meaning the spin of one or both objects is anti-aligned with respect to the orbital angular momentum of the system and this could be an indication of a non-isolated or dynamical formation channel of the binary. The added information from the effective precession spin parameter could also provide insight into dynamical formation channels. Figure ~\ref{fig:chi_p_pdf} plots the $\chi_\mathrm{p}$ posterior for all the events along with the parameter prior. We then compute the JS-divergence (JSD) between the prior and the individual posteriors, and GW190521 with a value of 0.181 shows the most deviation from the prior, as also seen in the figure. GW190403, GW191109, GW191225, GW200114 and GW170502 all have a JSD$>$0.1 between the $\chi_\mathrm{p}$ prior and respective posteriors. 



\begin{figure*}[t!]
    \centering
    \includegraphics[trim=100 10 0 0, width = 1.2\textwidth]{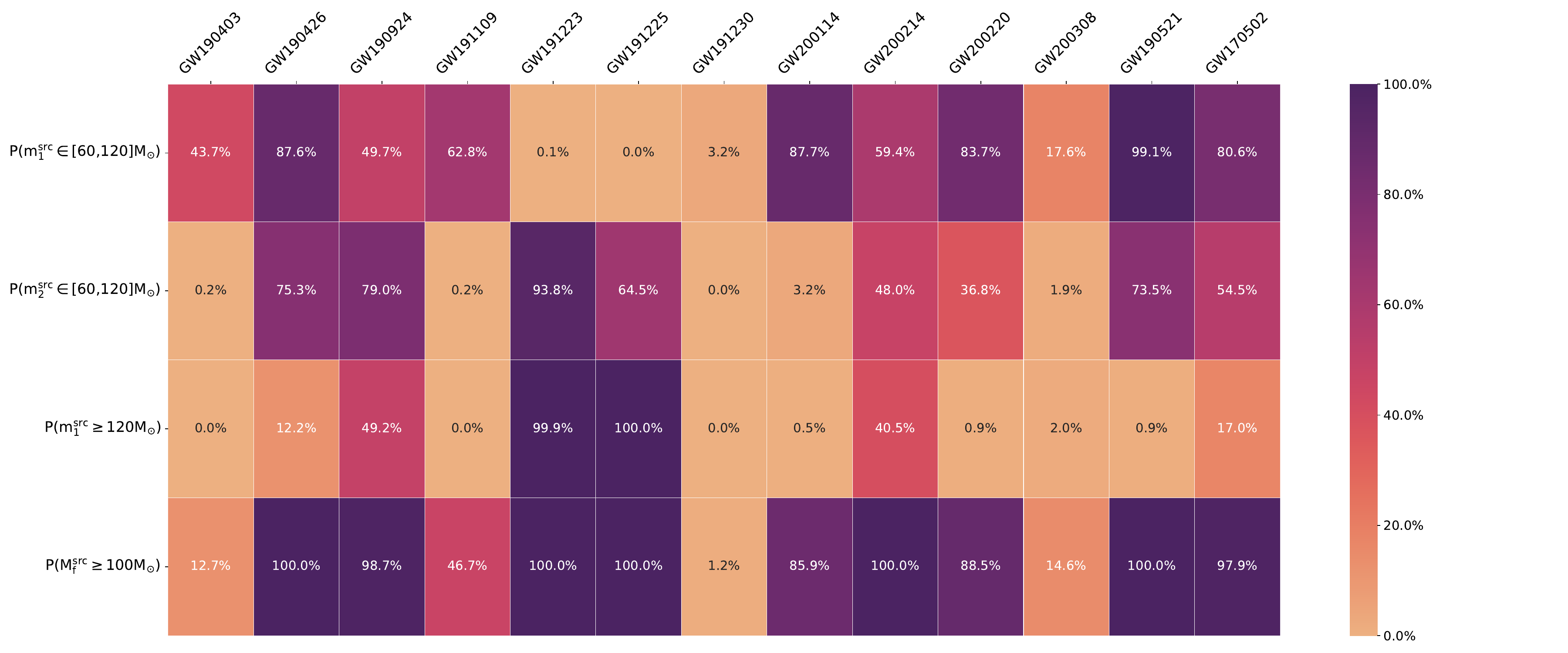}
    \caption{Heatmap showing the probabilities that the primary and secondary objects lie in the PISN mass-gap region of [60,120] $\mathrm{M}_{\odot}$ (first and second rows), the probability that the primary is heavier than 120 $\mathrm{M}_{\odot}$ (third row) and the probability that the final remnant is in the IMBH mass-region ($\geq 100\ \mathrm{M}_{\odot}$) (last row). The numbers in the individual rectangles denote the respective probabilities in percentages. The color bar shows the color scale of probabilities in percentages.}
    \label{figure:PISN_IMBH}
\end{figure*}

We also compute the probability of the component masses being in the pair-instability supernova (PISN) mass-gap of [60,120] $\mathrm{M}_{\odot}$, shown in Figure ~\ref{figure:PISN_IMBH}. All events except GW191230 and GW200308 have at least one component that falls in that gap with a probability greater than 40$\%$. GW191223 and GW191225 have more than 90$\%$ probability that the primary component has a mass $\geq 120 \ \mathrm{M}_{\odot}$. The existence of such black holes in the mass gap could also be evidence of their formation through hierarchical mergers of stellar black holes~\citep{Gerosa:2017kvu,Fishbach:2017dwv,Doctor:2019ruh,Yang:2019cbr} in nuclear star clusters~\citep{Antonini:2018auk,Baibhav:2020xdf} or dense gaseous disks of AGN~\citep{McKernan:2012rf,Bartos:2016dgn}. GW190426, GW191223, GW191225, GW200214 and GW190521 have 100$\%$ probability that the remnant black hole formed is in the intermediate-mass black hole region. GW191230 has the least probability at 1.2$\%$ of the remnant black hole being in the IMBH mass region.

\begin{figure*}
    \centering
    \includegraphics[trim=0 40 0 40, clip,width = 1.05\textwidth]{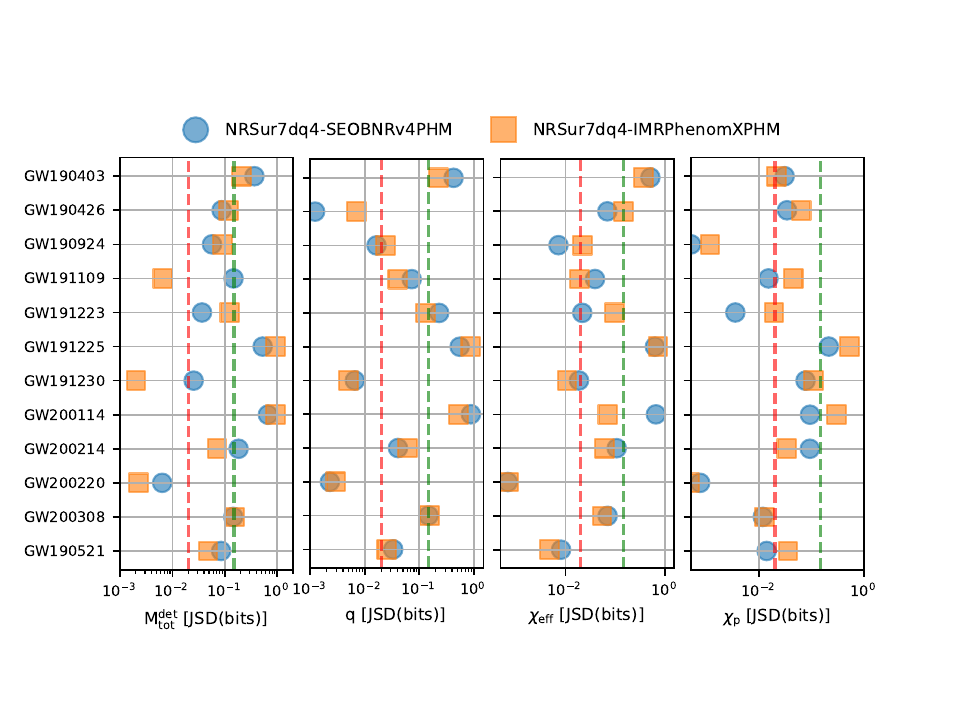}
    \caption{Jensen-Shannon divergence (JSD) values between the one-dimensional marginalized posteriors of total mass in detector-frame(M$_\mathrm{tot}^\mathrm{det}$), mass-ratio ($q$), the effective aligned spin parameter ($\chi_\mathrm{eff}$) and effective precession spin parameter($\chi_\mathrm{p}$). The dashed lines correspond to a JSD value of 0.02 bits (red) and 0.15 bits (green) respectively, which is a conventional value above which differences between posteriors are assumed to be significant. Note the GW190521 phenomenological model samples are from analysis using IMRPhenomPv3HM. }
    \label{figure:JSD_plot}
\end{figure*}
\section{Waveform Systematics} \label{sec:discussion}

 From Figure \ref{figure:violin plots}, the two events that stand out are GW191225 and GW200114, showing large differences between waveform families. The range of possible values for effective aligned spin is more constrained in comparison to the precession spin parameter. We see that drastic differences in inferred parameters between different waveform families seem to occur in the extremes of the parameter space, distinctly visible in the posterior differences of GW191223, GW191225, and GW200114.

We then quantify the differences between the waveform families using Jensen-Shannon divergences (JSD) and visualize it as in Figure \ref{figure:JSD_plot}, in particular we compute JSD between the posteriors of NRSur7dq4 and SEOBNRv4PHM (blue circle), and between NRSur7dq4 and IMRPhenomXPHM (orange square) for all the parameters shown in Fig. \ref{figure:violin plots}. A JSD cut-off of 0.02 ~\citep{Romero-Shaw:2020owr} (red dashed line) is used as a conservative number, and almost all of the events fall above this cut-off showing inconsistencies between inferred parameters, including the first confident high-mass BBH detection GW190521. In particular, looking at $\mathrm{M}^\mathrm{det}_\mathrm{Tot}$ and $q$, GW191109, GW191230 and GW200220 meet the JSD cut-off for at least two of the three waveform models. GW190924 and GW190521 lie just at the JSD cut-off boundary for $q$.  For the effective aligned spin constraints, GW190521 and GW20020 meet the requirement that the systematics are minute among considered waveform models, with GW190924 (meets for two waveform models), GW191109, GW191223 and GW191230 being just near the cut-off. With the spin-precession parameter, a varied behavior is observed among events; GW190924 and GW20020 fall well below the criteria whereas GW190403, GW191109, GW191223, GW200308, and GW190521 are close to this cut-off value.  

When using a less stringent cut-off of 0.15~\citep{Shaik:2019dym} (green dashed line), we see that more events appear to have consistent inferences among the waveform families. For the mass parameters, all but GW191225 and GW200114 fall under this relaxed cut-off, which is also notably visible in the posterior differences in Fig. \ref{figure:violin plots}. For $\chi_\mathrm{eff}$, GW190403 also shows systematic inconsistencies along with GW191225 and GW200114. GW191225 and GW200114 also are the only ones to fall above a JSD cut-off of 0.15 for the $\chi_\mathrm{p}$ parameter.

Some of the signals analyzed in this work seem to have instrumental noise artifacts, which is noted in the LVK O3 IMBH search results~\citep{LIGOScientific:2021tfm}.

\section{Conclusions}

We analyzed 11 lite IMBH signals in this study, and summarize the broad takeaways of this analysis below:

\begin{itemize}
    \item 5 of the 11 signals analysed have a probability $>$95$\%$ that the remnant is a black-hole in the IMBH region. 
    \item GW191223 is the heaviest black hole merger with an inferred source-frame total mass of 347$^{+71}_{-86}$ $\mathrm{M}_{\odot}$. 
    \item GW190403 is the most distant black hole merger observed with an inferred luminosity distance of 11.4$^{+7.6}_{-5.5}$ Gpc and an estimated source total mass of 87$^{+29}_{-21}$ $\mathrm{M}_{\odot}$.
    \item GW191225 is the closest heavy black-hole merger observed, at a luminosity distance 0.76$^{+0.49}_{-0.31}$ Gpc and a source total mass of 292$^{+33}_{-31}$ $\mathrm{M}_{\odot}$.
    \item GW190403 and GW200308 have the highest spin magnitude and GW191230 has the smallest spin magnitude for the system. 
    \item 5 of the 11 signals analysed have a probability $>$75$\%$ that at least one of the component masses fall in the PISN mass-gap region of [60,120] $\mathrm{M}_{\odot}$. 
    \item All lite IMBH candidates except GW191230 and GW200220 have waveform systematics worse than that exhibited by GW190521. Among these at least five events  (GW190924, GW191223, GW191225, GW200114 and GW200214) have potential data quality concerns from low-frequency noise~\citep{LIGOScientific:2021tfm}. 
\end{itemize}

Posterior samples from our analysis are publicly accessible~\citep{ruiz_rocha_2025_13872255}.

\section*{acknowledgements}

The authors would like to thank Imre Bartos for helpful feedback on the manuscript. We also thank Sergey Klimenko, Ryan Fisher, and Shubhanshu Tiwari for their useful comments. KJ and ABY acknowledge support from NSF PHY-2012057 grant and the Vanderbilt Scaling Grant from the Office of the Vice-Provost for
Research and Innovation. 
ROS gratefully acknowledges support from
NSF awards NSF PHY-1912632, PHY-2012057, PHY-2309172, AST- 2206321, and the Simons Foundation. This material is based upon work supported by NSF’s LIGO Laboratory which is a major facility fully funded by the National Science Foundation. This research has made use of data, software and/or web tools obtained from the Gravitational Wave Open Science Center (https://www.gw-openscience.org/ ), a service of LIGO Laboratory, the LIGO Scientific Collaboration and the Virgo Collaboration. LIGO Laboratory and Advanced LIGO are funded by the United States National Science Foundation (NSF) as well as the Science and Technology Facilities Council (STFC) of the United Kingdom, the Max-Planck-Society (MPS), and the State of Niedersachsen/Germany for support of the construction of Advanced LIGO and construction and operation of the GEO600 detector. Additional support for Advanced LIGO was provided by the Australian Research Council. Virgo is funded through the European Gravitational Observatory (EGO), by the French Centre National de Recherche Scientifique (CNRS), the Italian Istituto Nazionale di Fisica Nucleare (INFN), and the Dutch Nikhef, with contributions by institutions from Belgium, Germany, Greece, Hungary, Ireland, Japan, Monaco, Poland, Portugal, Spain. The authors are grateful for computational resources provided by the LIGO Laboratory and supported by National Science Foundation Grants PHY-0757058, PHY-0823459 and PHY-1626190, and IUCAA LDG cluster Sarathi.
KRR acknowledges and appreciates support from NSF NRT-2125764.


\bibliography{sample631}{}

\begin{thebibliography}{}
\expandafter\ifx\csname natexlab\endcsname\relax\def\natexlab#1{#1}\fi
\providecommand{\url}[1]{\href{#1}{#1}}
\providecommand{\dodoi}[1]{doi:~\href{http://doi.org/#1}{\nolinkurl{#1}}}
\providecommand{\doeprint}[1]{\href{http://ascl.net/#1}{\nolinkurl{http://ascl.net/#1}}}
\providecommand{\doarXiv}[1]{\href{https://arxiv.org/abs/#1}{\nolinkurl{https://arxiv.org/abs/#1}}}

\bibitem[{Aasi {et~al.}(2015)}]{LIGOScientific:2014pky}
Aasi, J., {et~al.} 2015, Class. Quant. Grav., 32, 074001, \dodoi{10.1088/0264-9381/32/7/074001}

\bibitem[{Abbott {et~al.}(2020{\natexlab{a}})}]{LIGOScientific:2020iuh}
Abbott, R., {et~al.} 2020{\natexlab{a}}, Phys. Rev. Lett., 125, 101102, \dodoi{10.1103/PhysRevLett.125.101102}

\bibitem[{Abbott {et~al.}(2020{\natexlab{b}})}]{LIGOScientific:2020ufj}
---. 2020{\natexlab{b}}, Astrophys. J. Lett., 900, L13, \dodoi{10.3847/2041-8213/aba493}

\bibitem[{Abbott {et~al.}(2022)}]{LIGOScientific:2021tfm}
---. 2022, Astron. Astrophys., 659, A84, \dodoi{10.1051/0004-6361/202141452}

\bibitem[{Abbott {et~al.}(2023)}]{KAGRA:2021vkt}
---. 2023, Phys. Rev. X, 13, 041039, \dodoi{10.1103/PhysRevX.13.041039}

\bibitem[{Abbott {et~al.}(2024)}]{LIGOScientific:2021usb}
---. 2024, Phys. Rev. D, 109, 022001, \dodoi{10.1103/PhysRevD.109.022001}

\bibitem[{Acernese {et~al.}(2015)}]{VIRGO:2014yos}
Acernese, F., {et~al.} 2015, Class. Quant. Grav., 32, 024001, \dodoi{10.1088/0264-9381/32/2/024001}

\bibitem[{Ade {et~al.}(2016)Ade, Aghanim, Arnaud, Ashdown, Aumont, Baccigalupi, Banday, Barreiro, Bartlett, Bartolo, Battaner, Battye, Benabed, Benoît, Benoit-Lévy, Bernard, Bersanelli, Bielewicz, Bock, Bonaldi, Bonavera, Bond, Borrill, Bouchet, Boulanger, Bucher, Burigana, Butler, Calabrese, Cardoso, Catalano, Challinor, Chamballu, Chary, Chiang, Chluba, Christensen, Church, Clements, Colombi, Colombo, Combet, Coulais, Crill, Curto, Cuttaia, Danese, Davies, Davis, de~Bernardis, de~Rosa, de~Zotti, Delabrouille, Désert, Di~Valentino, Dickinson, Diego, Dolag, Dole, Donzelli, Doré, Douspis, Ducout, Dunkley, Dupac, Efstathiou, Elsner, Enßlin, Eriksen, Farhang, Fergusson, Finelli, Forni, Frailis, Fraisse, Franceschi, Frejsel, Galeotta, Galli, Ganga, Gauthier, Gerbino, Ghosh, Giard, Giraud-Héraud, Giusarma, Gjerløw, González-Nuevo, Górski, Gratton, Gregorio, Gruppuso, Gudmundsson, Hamann, Hansen, Hanson, Harrison, Helou, Henrot-Versillé, Hernández-Monteagudo, Herranz, Hildebrandt, Hivon, Hobson, Holmes,
  Hornstrup, Hovest, Huang, Huffenberger, Hurier, Jaffe, Jaffe, Jones, Juvela, Keihänen, Keskitalo, Kisner, Kneissl, Knoche, Knox, Kunz, Kurki-Suonio, Lagache, Lähteenmäki, Lamarre, Lasenby, Lattanzi, Lawrence, Leahy, Leonardi, Lesgourgues, Levrier, Lewis, Liguori, Lilje, Linden-Vørnle, López-Caniego, Lubin, Macías-Pérez, Maggio, Maino, Mandolesi, Mangilli, Marchini, Maris, Martin, Martinelli, Martínez-González, Masi, Matarrese, McGehee, Meinhold, Melchiorri, Melin, Mendes, Mennella, Migliaccio, Millea, Mitra, Miville-Deschênes, Moneti, Montier, Morgante, Mortlock, Moss, Munshi, Murphy, Naselsky, Nati, Natoli, Netterfield, Nørgaard-Nielsen, Noviello, Novikov, Novikov, Oxborrow, Paci, Pagano, Pajot, Paladini, Paoletti, Partridge, Pasian, Patanchon, Pearson, Perdereau, Perotto, Perrotta, Pettorino, Piacentini, Piat, Pierpaoli, Pietrobon, Plaszczynski, Pointecouteau, Polenta, Popa, Pratt, Prézeau, Prunet, Puget, Rachen, Reach, Rebolo, Reinecke, Remazeilles, Renault, Renzi, Ristorcelli, Rocha, Rosset,
  Rossetti, Roudier, Rouillé~d’Orfeuil, Rowan-Robinson, Rubiño-Martín, Rusholme, Said, Salvatelli, Salvati, Sandri, Santos, Savelainen, Savini, Scott, Seiffert, Serra, Shellard, Spencer, Spinelli, Stolyarov, Stompor, Sudiwala, Sunyaev, Sutton, Suur-Uski, Sygnet, Tauber, Terenzi, Toffolatti, Tomasi, Tristram, Trombetti, Tucci, Tuovinen, Türler, Umana, Valenziano, Valiviita, Van~Tent, Vielva, Villa, Wade, Wandelt, Wehus, White, White, Wilkinson, Yvon, Zacchei, \& Zonca}]{Placnk15}
Ade, P. A.~R., Aghanim, N., Arnaud, M., {et~al.} 2016, Astronomy \& Astrophysics, 594, A13, \dodoi{10.1051/0004-6361/201525830}

\bibitem[{Akutsu {et~al.}(2021)}]{KAGRA:2020tym}
Akutsu, T., {et~al.} 2021, PTEP, 2021, 05A101, \dodoi{10.1093/ptep/ptaa125}

\bibitem[{Antonini {et~al.}(2019)Antonini, Gieles, \& Gualandris}]{Antonini:2018auk}
Antonini, F., Gieles, M., \& Gualandris, A. 2019, Mon. Not. Roy. Astron. Soc., 486, 5008, \dodoi{10.1093/mnras/stz1149}

\bibitem[{Aubin {et~al.}(2021)}]{Aubin:2020goo}
Aubin, F., {et~al.} 2021, Class. Quant. Grav., 38, 095004, \dodoi{10.1088/1361-6382/abe913}

\bibitem[{Baibhav {et~al.}(2020)Baibhav, Gerosa, Berti, Wong, Helfer, \& Mould}]{Baibhav:2020xdf}
Baibhav, V., Gerosa, D., Berti, E., {et~al.} 2020, Phys. Rev. D, 102, 043002, \dodoi{10.1103/PhysRevD.102.043002}

\bibitem[{Bartos {et~al.}(2017)Bartos, Kocsis, Haiman, \& M\'arka}]{Bartos:2016dgn}
Bartos, I., Kocsis, B., Haiman, Z., \& M\'arka, S. 2017, Astrophys. J., 835, 165, \dodoi{10.3847/1538-4357/835/2/165}

\bibitem[{{Cornish} \& {Littenberg}(2015)}]{2015CQGra..32m5012C}
{Cornish}, N.~J., \& {Littenberg}, T.~B. 2015, Classical and Quantum Gravity, 32, 135012, \dodoi{10.1088/0264-9381/32/13/135012}

\bibitem[{Cornish {et~al.}(2021)Cornish, Littenberg, B\'ecsy, Chatziioannou, Clark, Ghonge, \& Millhouse}]{Cornish:2020dwh}
Cornish, N.~J., Littenberg, T.~B., B\'ecsy, B., {et~al.} 2021, Phys. Rev. D, 103, 044006, \dodoi{10.1103/PhysRevD.103.044006}

\bibitem[{Dal~Canton {et~al.}(2021)Dal~Canton, Nitz, Gadre, Cabourn~Davies, Villa-Ortega, Dent, Harry, \& Xiao}]{DalCanton:2020vpm}
Dal~Canton, T., Nitz, A.~H., Gadre, B., {et~al.} 2021, Astrophys. J., 923, 254, \dodoi{10.3847/1538-4357/ac2f9a}

\bibitem[{Doctor {et~al.}(2019)Doctor, Wysocki, O'Shaughnessy, Holz, \& Farr}]{Doctor:2019ruh}
Doctor, Z., Wysocki, D., O'Shaughnessy, R., Holz, D.~E., \& Farr, B. 2019, \dodoi{10.3847/1538-4357/ab7fac}

\bibitem[{Ewing {et~al.}(2024)}]{Ewing:2023qqe}
Ewing, B., {et~al.} 2024, Phys. Rev. D, 109, 042008, \dodoi{10.1103/PhysRevD.109.042008}

\bibitem[{Farag {et~al.}(2022)Farag, Renzo, Farmer, Chidester, \& Timmes}]{Farag:2022jcc}
Farag, E., Renzo, M., Farmer, R., Chidester, M.~T., \& Timmes, F.~X. 2022, Astrophys. J., 937, 112, \dodoi{10.3847/1538-4357/ac8b83}

\bibitem[{Farmer {et~al.}(2019)Farmer, Renzo, de~Mink, Marchant, \& Justham}]{Farmer:2019jed}
Farmer, R., Renzo, M., de~Mink, S.~E., Marchant, P., \& Justham, S. 2019, \dodoi{10.3847/1538-4357/ab518b}

\bibitem[{Fishbach \& Holz(2020)}]{Fishbach:2020qag}
Fishbach, M., \& Holz, D.~E. 2020, Astrophys. J. Lett., 904, L26, \dodoi{10.3847/2041-8213/abc827}

\bibitem[{Fishbach {et~al.}(2017)Fishbach, Holz, \& Farr}]{Fishbach:2017dwv}
Fishbach, M., Holz, D.~E., \& Farr, B. 2017, Astrophys. J. Lett., 840, L24, \dodoi{10.3847/2041-8213/aa7045}

\bibitem[{Gerosa \& Berti(2017)}]{Gerosa:2017kvu}
Gerosa, D., \& Berti, E. 2017, Phys. Rev. D, 95, 124046, \dodoi{10.1103/PhysRevD.95.124046}

\bibitem[{Greene {et~al.}(2020)Greene, Strader, \& Ho}]{Greene_2020}
Greene, J.~E., Strader, J., \& Ho, L.~C. 2020, Annual Review of Astronomy and Astrophysics, 58, 257, \dodoi{10.1146/annurev-astro-032620-021835}

\bibitem[{Healy \& Lousto(2017)}]{Healy:2016lce}
Healy, J., \& Lousto, C.~O. 2017, Phys. Rev. D, 95, 024037, \dodoi{10.1103/PhysRevD.95.024037}

\bibitem[{Hofmann {et~al.}(2016)Hofmann, Barausse, \& Rezzolla}]{Hofmann:2016yih}
Hofmann, F., Barausse, E., \& Rezzolla, L. 2016, Astrophys. J. Lett., 825, L19, \dodoi{10.3847/2041-8205/825/2/L19}

\bibitem[{Hoy \& Raymond(2021)}]{Hoy:2020vys}
Hoy, C., \& Raymond, V. 2021, SoftwareX, 15, 100765, \dodoi{10.1016/j.softx.2021.100765}

\bibitem[{{Jani} {et~al.}(2020){Jani}, {Shoemaker}, \& {Cutler}}]{2020NatAs...4..260J}
{Jani}, K., {Shoemaker}, D., \& {Cutler}, C. 2020, Nature Astronomy, 4, 260, \dodoi{10.1038/s41550-019-0932-7}

\bibitem[{Keitel {et~al.}(2017)}]{Keitel:2016krm}
Keitel, D., {et~al.} 2017, Phys. Rev. D, 96, 024006, \dodoi{10.1103/PhysRevD.96.024006}

\bibitem[{Klimenko {et~al.}(2016)}]{Klimenko:2015ypf}
Klimenko, S., {et~al.} 2016, Phys. Rev. D, 93, 042004, \dodoi{10.1103/PhysRevD.93.042004}

\bibitem[{{Lange} {et~al.}(2018){Lange}, {O'Shaughnessy}, \& {Rizzo}}]{gwastro-PENR-RIFT}
{Lange}, J., {O'Shaughnessy}, R., \& {Rizzo}, M. 2018, Submitted to PRD; available at arxiv:1805.10457

\bibitem[{Littenberg \& Cornish(2015)}]{PhysRevD.91.084034}
Littenberg, T.~B., \& Cornish, N.~J. 2015, Phys. Rev. D, 91, 084034, \dodoi{10.1103/PhysRevD.91.084034}

\bibitem[{Marchant \& Moriya(2020)}]{Marchant:2020haw}
Marchant, P., \& Moriya, T. 2020, Astron. Astrophys., 640, L18, \dodoi{10.1051/0004-6361/202038902}

\bibitem[{McKernan {et~al.}(2012)McKernan, Ford, Lyra, \& Perets}]{McKernan:2012rf}
McKernan, B., Ford, K. E.~S., Lyra, W., \& Perets, H.~B. 2012, Mon. Not. Roy. Astron. Soc., 425, 460, \dodoi{10.1111/j.1365-2966.2012.21486.x}

\bibitem[{Mishra {et~al.}(2025)Mishra, Bhaumik, Gayathri, Szczepa\'nczyk, Bartos, \& Klimenko}]{Mishra:2024zzs}
Mishra, T., Bhaumik, S., Gayathri, V., {et~al.} 2025, Phys. Rev. D, 111, 023054, \dodoi{10.1103/PhysRevD.111.023054}

\bibitem[{Mishra {et~al.}(2022)}]{Mishra:2022ott}
Mishra, T., {et~al.} 2022, Phys. Rev. D, 105, 083018, \dodoi{10.1103/PhysRevD.105.083018}

\bibitem[{Nitz {et~al.}(2023)Nitz, Kumar, Wang, Kastha, Wu, Sch\"afer, Dhurkunde, \& Capano}]{Nitz:2021zwj}
Nitz, A.~H., Kumar, S., Wang, Y.-F., {et~al.} 2023, Astrophys. J., 946, 59, \dodoi{10.3847/1538-4357/aca591}

\bibitem[{{O'Shaughnessy} {et~al.}(2017){O'Shaughnessy}, {Blackman}, \& {Field}}]{gwastro-PE-AlternativeArchitecturesROM}
{O'Shaughnessy}, R., {Blackman}, J., \& {Field}, S. 2017, Class. Quant. Grav., \dodoi{10.1088/1361-6382/aa7649}

\bibitem[{Ossokine {et~al.}(2020)}]{Ossokine:2020kjp}
Ossokine, S., {et~al.} 2020, Phys. Rev. D, 102, 044055, \dodoi{10.1103/PhysRevD.102.044055}

\bibitem[{{Pankow} {et~al.}(2015){Pankow}, {Brady}, {Ochsner}, \& {O'Shaughnessy}}]{gwastro-PE-AlternativeArchitectures}
{Pankow}, C., {Brady}, P., {Ochsner}, E., \& {O'Shaughnessy}, R. 2015, \prd, 92, 023002, \dodoi{10.1103/PhysRevD.92.023002}

\bibitem[{Pratten {et~al.}(2021)}]{Pratten:2020ceb}
Pratten, G., {et~al.} 2021, Phys. Rev. D, 103, 104056, \dodoi{10.1103/PhysRevD.103.104056}

\bibitem[{Romero-Shaw {et~al.}(2020)}]{Romero-Shaw:2020owr}
Romero-Shaw, I.~M., {et~al.} 2020, Mon. Not. Roy. Astron. Soc., 499, 3295, \dodoi{10.1093/mnras/staa2850}

\bibitem[{Ruiz-Rocha \& Yelikar(2025)}]{ruiz_rocha_2025_13872255}
Ruiz-Rocha, K., \& Yelikar, A. 2025, Properties of 'Lite' Intermediate-Mass Black Hole Candidates in LIGO-Virgo's Third Observing Run,  Zenodo, \dodoi{10.5281/zenodo.13872255}

\bibitem[{Shaik {et~al.}(2020)Shaik, Lange, Field, O'Shaughnessy, Varma, Kidder, Pfeiffer, \& Wysocki}]{Shaik:2019dym}
Shaik, F.~H., Lange, J., Field, S.~E., {et~al.} 2020, Phys. Rev. D, 101, 124054, \dodoi{10.1103/PhysRevD.101.124054}

\bibitem[{Tsukada {et~al.}(2023)}]{Tsukada:2023edh}
Tsukada, L., {et~al.} 2023, Phys. Rev. D, 108, 043004, \dodoi{10.1103/PhysRevD.108.043004}

\bibitem[{Udall {et~al.}(2020)Udall, Jani, Lange, O'Shaughnessy, Clark, Cadonati, Shoemaker, \& Holley-Bockelmann}]{Udall:2019wtd}
Udall, R., Jani, K., Lange, J., {et~al.} 2020, Astrophys. J., 900, 80, \dodoi{10.3847/1538-4357/abab9d}

\bibitem[{Varma {et~al.}(2019)Varma, Field, Scheel, Blackman, Gerosa, Stein, Kidder, \& Pfeiffer}]{Varma:2019csw}
Varma, V., Field, S.~E., Scheel, M.~A., {et~al.} 2019, Phys. Rev. Research., 1, 033015, \dodoi{10.1103/PhysRevResearch.1.033015}

\bibitem[{{Wofford} {et~al.}(2023){Wofford}, {Yelikar}, {Gallagher}, {Champion}, {Wysocki}, {Delfavero}, {Lange}, {Rose}, {Valsan}, {Morisaki}, {Read}, {Henshaw}, \& {O'Shaughnessy}}]{gwastro-RIFT-Update}
{Wofford}, J., {Yelikar}, A.~B., {Gallagher}, H., {et~al.} 2023, \prd, 107, 024040, \dodoi{10.1103/PhysRevD.107.024040}

\bibitem[{Woosley \& Heger(2021)}]{Woosley:2021xba}
Woosley, S.~E., \& Heger, A. 2021, Astrophys. J. Lett., 912, L31, \dodoi{10.3847/2041-8213/abf2c4}

\bibitem[{{Wysocki} {et~al.}(2019){Wysocki}, {O'Shaughnessy}, {Lange}, \& {Fang}}]{gwastro-PENR-RIFT-GPU}
{Wysocki}, D., {O'Shaughnessy}, R., {Lange}, J., \& {Fang}, Y.-L.~L. 2019, \prd, 99, 084026, \dodoi{10.1103/PhysRevD.99.084026}

\bibitem[{Yang {et~al.}(2019)}]{Yang:2019cbr}
Yang, Y., {et~al.} 2019, Phys. Rev. Lett., 123, 181101, \dodoi{10.1103/PhysRevLett.123.181101}

\end{thebibliography}
\bibliographystyle{aasjournal}

\end{document}